# Influence of center distance on high harmonic generation

P.A. Golovinski [1,2], A.A. Drobyshev [2]

[1] *Moscow Institute of Physics and Technology (State University), 141700, Moscow, Russia*
[2] *Voronezh State University of Architecture and Civil Engineering, 394006, Voronezh, Russia*

The nonlinear ionization with accompanied photorecombination on closely located center has been considered. The radiation spectrum has been calculated in the frame of the Lewenstein-Corkum approach. The dependence of radiation on the distance between ionization point and the recombination center, and the frequency and laser field strength has been investigated.

Key words: high harmonic, laser field, electron, frequency, two-centers, photorecombination.

## 1. Introduction

For physical description of photorecombination of an electron on the parent atom, one has a approach [1], assuming the radiation is caused by an active electron, that is ejected from atom and moving in the laser field. The act of photon emission occurs when electron comes back to the atom, where it recombines to the ground state. Characteristic frequency of the emitted photon in the process is proportional to the ponderomotive potential and quadratic with respect to the electric field strength. In superatomic laser fields, electron radiation can be calculated as a result of interaction of electron wave packet with atomic core [2]. To this limit, spectrum tends to be continuum [3, 4], that is very suitable to the attosecond pulse formation.

One of the most known approximation for the high harmonic generation is based on the short distance potential, in particular in the Floquet or quasienergy consideration. The essential achievement of this approach is completely analytical description of the high harmonic generation and justification of the three step model of the process [5, 6].

High harmonic generation in many-atomic systems demonstrates new peculiarities including general increasing of the yield signal [7, 8], modification of frequency dependence [9], and increasing of the cutoff frequency in the harmonic spectrum [10]. The feature of the high harmonic generation in atomic gases is only odd numbers of harmonics in the spectrum that is a result of the mirror symmetry of atoms [11]. For the systems without the mirror symmetry all numbers of harmonics are presented in the spectrum.

The photorecombination on the closely placed centers tends to have some new characteristics, for example, it is non quadratic dependence of the cutoff frequency on the laser field strength. At the same time, this process has just the same nature as the high harmonic generation on the individual atom. It is explained by the electron wave packet formation and its propagation with transitions to the ground state of a system with photon emission. In the [12] this process was calculated in quasi static limit for electric field. In the present paper, the problem of photorecombination on the closely located center is solved for nonlinear ionization in harmonic laser field.

## 2. Perturbation for unsteady states

The generic perturbation theory requires some modifications in view of strong interaction with a laser field. Adequate machinery has been developed in the nonequilibrium statistical physics [12]. The total system Hamiltonian is a sum of Hamiltonian $H(t)$ for electron in strong laser field and classical potential of centers, and perturbation $H_1(t)$ that is operator of spontaneous emission of photon with frequency $\Omega$. We follow the formal perturbation theory [14], and the Srödinger equation in the interaction picture takes the form

$$i\frac{\partial \Phi}{\partial t} = \left(S(t,t_0)^{-1} H_1(t) S(t,t_0)\right)\Phi, \quad (1)$$

where $S(t,t_0)$ is the time evolution operator for an electron in the external electromagnetic field. In the first order of the perturbation theory with respect to the interaction $H_1(t)$, the solution of the Eq. (1) one may write down as

$$\Phi(t) = -i\int_{t_0}^{t} \left(S(\tau,t_0)^{-1} H_1(\tau) S(\tau,t_0)\right)\Phi(t_0)\, d\tau + \Phi(t_0). \quad (2)$$

For large times $t \to \pm\infty$, there is no external laser field, and the system comes to one of the initial or final steady states.

The operator of photon emission with polarization $\mathbf{e}_\alpha$ in the quantization volume $V$ in the dipole approximation has the form [15]

$$H_1(t) = w\exp(i\Omega t),\; w = \left(\frac{2\pi}{V\Omega}\right)^{1/2}(\mathbf{e}_\alpha \hat{\mathbf{p}}). \quad (3)$$

Let us use the Lewenstein-Corkum approach, that contains three assumptions [16]: in initial time the system has only one bound electron state, the electron motion in continuum is described taking into account the laser field, the initial bound state is slightly distorted by the laser field. We take the velocity gauge for interaction of electron with laser field. This choice has some analytical advantages in comparison with the length gauge [16], and it supports a higher accuracy for the Lewenstein-Corkum approach [17].

Electron wave functions $\Phi_l$ in initial and final states ($l = i, f$) we represent as a superposition of discrete and continuum states

$$\Phi_l = \exp(-iE_l t)\left(a_l(t)|k\rangle + \int d\mathbf{p}\, b_{\mathbf{p}l}(t)|\mathbf{p}\rangle\right), \quad (4)$$

and $a_l(t) \approx 1$ according to the low intensity assumption. From Eq. (4) and Eq. (1), it follows the projections on the continuum states obey the differential equation for coefficients $b_{\mathbf{p}l}(t)$. Its solution has the form

$$b_{\mathbf{p}l}(t) = -i\int_{-\infty}^{t} \mathbf{k}(\tau)\mathbf{d}_{\mathbf{p}l} \exp\left(-i\int_{\tau}^{t} g_l(\tau_1)\, d\tau_1\right) d\tau, \quad (5)$$

where $g_l(\tau) = (\mathbf{p} + \mathbf{k}(\tau))^2/2 - E_l$. An initial condition $b_{\mathbf{p}l}(0) = 0$ fulfils automatically.

The probability amplitude is

$$M_{fi}(t) = -i\int_{-\infty}^{t} \exp(i(E_f - E_i + \Omega)\tau) B_{fi}(\tau)\, d\tau, \quad (6)$$

where

$$B_{fi}(\tau) = \int d\mathbf{p} \langle f|w|\mathbf{p}\rangle b_{\mathbf{p}i}(\tau) = \left(\frac{2\pi}{V\Omega}\right)^{1/2} \mathbf{e}_\alpha \mathbf{P}_{fi}(\tau), \quad \mathbf{P}_{fi}(\tau) = \int d\mathbf{p}\, \mathbf{d}_{f\mathbf{p}} b_{\mathbf{p}i}(\tau) \qquad (7)$$

For monochromatic field [18] one has

$$B_{fi}(\tau) = \sum_n \tilde{B}_{fi}(\Omega_n)\exp(-i\Omega_n\tau), \quad \tilde{B}_{fi}(\Omega_n) = \left(\frac{2\pi}{V\Omega}\right)^{1/2} \mathbf{e}_\alpha \tilde{\mathbf{P}}_{fi}(\Omega_n). \qquad (8)$$

Here $\tilde{\mathbf{P}}(\Omega_n)$ is Fourier transform of the polarization vector $\mathbf{P}_{fi}(\tau)$.

The transition probability per unit of time

$$|M_{fi}|^2 / t = 2\pi |\tilde{B}_{fi}(\Omega_n)|^2 \delta(E_f - E_i + \Omega - \Omega_n). \qquad (9)$$

In the Eq. (9) the energy conservation is fixed by the equality $E = E_i + \Omega_n$, when final state has energy $E_f$. For a certainty, we assume $E_f = E_i = -I_p$, and $\Omega = \Omega_n$.

### 3. Photorecombination

The intensity of spontaneous photon emission with a frequency $\Omega_n$ is then

$$I_n = \frac{4\Omega_n^2}{3c^3} |\tilde{\mathbf{P}}_{fi}(\Omega_n)|^2. \qquad (10)$$

The initial state we take in the form of Gaussian wave function [12]

$$\langle \mathbf{r}|l\rangle = \left(\frac{\alpha}{\pi}\right)^{3/4} \exp\left(-\frac{\alpha r^2}{2}\right), \qquad (11)$$

where $\alpha \sim I_p$. This simple form is very convenient for analytical integration in the Eq. (7). Any more complex wave function can be represented as superposition of Gaussian type functions (radial basis functions). After integration we get

$$\mathbf{d}_{\mathbf{p}i} = i\mathbf{p}\left(\frac{1}{\pi\alpha}\right)^{3/4}\exp\left(-\frac{p^2}{2\alpha}\right), \quad \mathbf{d}_{f\mathbf{p}} = \mathbf{d}_{\mathbf{p}i}\exp(i\mathbf{pR}), \qquad (12)$$

where $\mathbf{R}$ is a vector of the recombination center position, $\mathbf{pR}$ is the phase shift after propagation electron wave between centers. For polarization vector one can obtain

$$\mathbf{P}_{fi}(t) = -i\left(\frac{1}{\alpha}\right)^{3/2} \int_{-\infty}^{t} \exp\left(-i(\varphi + I_p\Delta t - \mathbf{a}_1^2 \alpha\Delta t/8\beta)\right) \times$$

$$\times \left(\frac{1 - i\alpha\Delta t/2}{\beta}\right)^{7/2} \mathbf{a}_1(\mathbf{ka}_1)\exp\left(-\frac{\mathbf{a}_1^2}{4\beta}\right) d\tau. \qquad (13)$$

We introduce new definitions: $\mathbf{a}_1 = \mathbf{a} - \mathbf{R}$, $\beta = 1/\alpha + \alpha \Delta t^2/4$, $\Delta t = t - \tau$, $\mathbf{a}(t,\tau) = \int_\tau^t \mathbf{k}(\tau) d\tau$, $\varphi(t,\tau) = \int_\tau^t k^2(\tau)/2 \, d\tau$. Function $\mathbf{a}_1(t)$ is classical electron displacement with respect to recombination center under the action of the laser field.

For quantum dots taken as "artificial atoms" [19] the free electron mass is changed by effective electron mass in all equations.

## 4. Electron motion after ionization

In the low frequency limit, an electron motion under the action of the force $F(t) = -F_0 \cos(\omega t)$ is essentially classical. We denote $\tau = \omega t$, $u = x\omega^2/F_0$, where $x$ is the distance between the centers of ionization and recombination. Then, the dependence of the scaled coordinate $u$ as a function of time has the form

$$u(\tau) = -\cos(\tau + \varphi_0) - \tau \sin \varphi_0 + \cos \varphi_0. \qquad (14)$$

The Eq. (14) contains two essential terms, a periodical oscillation and a drift with a constant velocity $-\sin \varphi_0$. Peculiarities of the electron motion depend on the value of the initial phase $\varphi_0$. High harmonic generation on the parent atom is caused comeback electrons with phases $\varphi_0 \in (-\pi, -\pi/2) \cup (0, \pi/2)$. In contrast, all electrons with phases $\varphi_0 \in (-\pi, \pi/2)$ takes a part in the recombination on a secondary center.

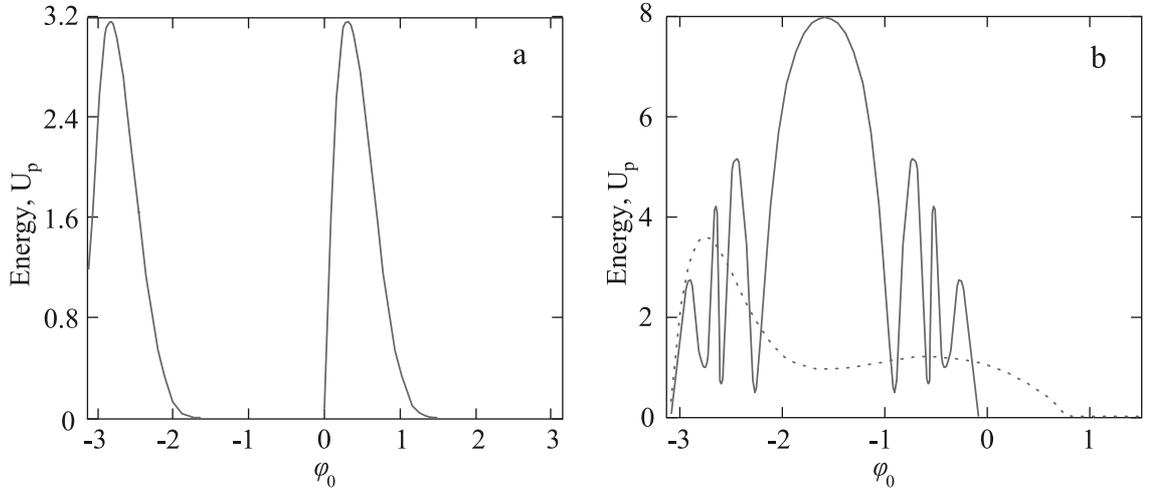

**Fig. 1.** Electron energy dependence on the initial laser phase at the point of comeback to the initial ion (a), and at the second recombination center (b).

Fig. 1 shows the dependence of the electron energy as a function of the initial phase. In the Fig. 1a such curve is plotted for the time of comeback to the ionization center, the Fig. 1b demonstrates this dependence for a secondary center recombination and for distances between centers $R = 95$ a.u. (solid line), $R = 5$ a.u. (points). All calculations are attributed to the first time of passing the center of recombination. The energy is measured in the units of ponderomotive potential $U_p = F_0^2/4\omega^2$. For the high harmonic generation on the initial center the maximum kinetic energy is $E_{\max} \approx 3.2 U_p$, accordingly, maximum harmonic frequency is $\Omega_{max} \approx I_p + 3.2U_p$ [1]. The maximum radiation frequency for recombination on the secondary center depends on the distance between the center of ionization and the center of recombination. The maximum value of the kinetic energy is $8U_p$, that is accessible for $\varphi_0 = -\pi/2$ at the moments of time $\tau_n = \pi(1+2n)$, where $n = 0, 1, 2 ...$, and corresponding distances are $R_n = \tau_n F_0/\omega^2$.

## 5. Recombination spectrum

The comparison of the calculation, that was produced according to Eq. (10) and Eq. (13), with experimental data [20] demonstrates that our approach reproduces the results of high harmonic generations in atomic gases [16].

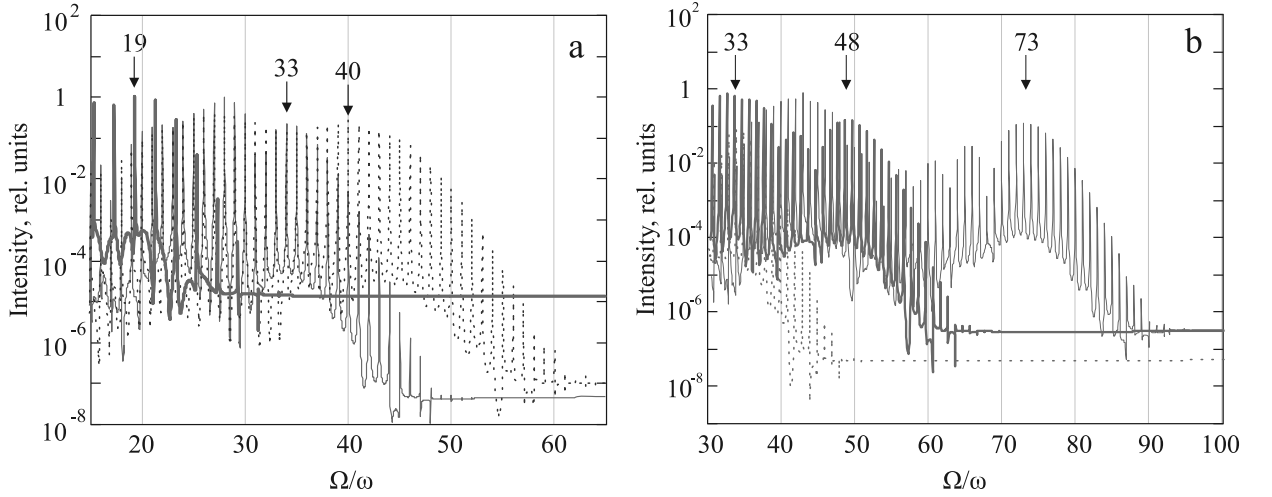

**Fig. 2.** Spectral distribution of the recombination radiation for the different distances between centers (a), and for the different wavelengths of the laser field (b). Left side (a): solid line shows the recombination on the ionization center, thin line corresponds to the distance 30 a.u., points correspond to the distance 55 a.u.. Right side (b): points correspond to $\lambda = 800$ nm, solid line corresponds to $\lambda = 1000$ nm, thin line corresponds to $\lambda = 1200$ nm.

Fig. 2a shows the spectrum of radiation for different distances between centers having the same ionization potential $I_p = 10$ eV and for the laser field intensity $I = 10^{14}$ W/cm$^2$. The spectrum is shifted to the blue region, when the distance between centers increases. For the case of a static electric field [12], the classical energy $E_{max}$ of the recombined electron increases proportional to the distance $R$. The cutoff frequency is function of coordinate also, and $\Omega_{max} = I_p + E_{max}$. In contrast to that, in the two-center case the characteristic number of harmonic, calculated with the help of the Eq. (14), is $\Omega/\omega = 19$ for $R = 0$ a.u., $\Omega/\omega = 33$ for $R = 30$ a.u., and $\Omega/\omega = 40$ for $R = 55$ a.u..

Fig 2b demonstrates the results of numerical simulations of radiation spectrum for different laser wavelengths and fixed distance $R = 30$ a.u. when $I_p = 10$ eV and $I = 10^{14}$ W/cm$^2$. The cutoff frequencies are depicted. The classical estimation of the harmonic number at the boundary before decreasing intensity for the laser field with a wavelength $\lambda = 800$ nm gives $\Omega/\omega = 33$, for $\lambda = 1000$ nm it is $\Omega/\omega = 48$, and for $\lambda = 1200$ nm it is $\Omega/\omega = 73$.

## 6. Summary

We described photorecombination of electron after the nonlinear ionization in a monochromatic laser field. Following the Lewenstein-Corkum approach we obtained analytical expression for matrix element of spontaneous photon emission by an active electron moving in a strong optical field. The radiation spectrum for the process of photorecombination on a secondary center has been calculated. This new type of spectrum has all numbers of harmonics in spectrum, that is differ of the harmonic generation on a parent center, where only odd harmonics are presented. The spectrum is shifted to a blue side of wavelength with respect to the same for a parent atom. The maximum frequency in the spectrum of recombination on the secondary center is determined by the maximum kinetic energy of an electron, and can reach eight ponderomotive potentials. Our consideration proves that two and, perhaps, many-center systems are more effective as objects for high harmonic generation comparing with atom gases.

This work was supported by the RFFI (grant №13-07-00270) and the Ministry of Education and Science of the Russian Federation (№ 2014.19-2881).